\documentclass[12]{article}
\usepackage[english]{babel}
\usepackage[a4paper, total={5.5in, 9in}]{geometry}
\usepackage{graphicx} 
\usepackage{booktabs}
\usepackage{amsmath}
\usepackage[hidelinks]{hyperref}
\usepackage{microtype}
\usepackage{amssymb}
\usepackage{natbib}
\usepackage{xcolor}
\usepackage{authblk} 
\usepackage[font={small,it}]{caption}
\usepackage{subcaption}
\usepackage{bbm}
\usepackage[normalem]{ulem}

\providecommand{\keywords}[1]{\textbf{Keywords:} #1}

\title{Selective inference for fMRI cluster-wise analysis, issues, and recommendations for critical vector selection: A comment on Blain et al.}
\author[1]{Angela Andreella \thanks{\textbf{Corresponding author}: angela.andreella@unive.it}}
\author[2]{Anna Vesely}
\author[3]{Wouter Weeda}
\author[4]{Jelle Goeman}

\affil[1]{\footnotesize Department of Economics, Ca' Foscari University of Venice, Venice, Italy}
\affil[2]{\footnotesize Department of Statistical Sciences, University of Bologna, Bologna, Italy}
\affil[3]{\footnotesize Department of Psychology, Leiden University, Leiden, The Netherlands}
\affil[4]{\footnotesize Department of Biomedical Data Sciences, Leiden University Medical Center, Leiden, The Netherlands}
\date{}

\begin{document}

\newpage

\maketitle

\begin{abstract}
Two permutation-based methods for simultaneous inference on the proportion of active voxels in cluster-wise brain imaging analysis have recently been published: Notip \citep{blain2022notip} and pARI \citep{andreella2023permutation}. Both rely on the definition of a critical vector of ordered $p$-values, chosen from a family of candidate vectors, but differ in how the family is defined: computed from randomization of external data for Notip and determined a priori for pARI. These procedures were compared to other proposals in the literature, but an extensive comparison between the two methods is missing due to their parallel publication. We provide such a comparison and find that pARI outperforms Notip if both methods are applied under their recommended settings. However, each method carries different advantages and drawbacks.
\end{abstract}


\section{Introduction}

Cluster-extent-based thresholding is a common approach in functional Magnetic Resonance Imaging (fMRI) analysis to explore which parts of the human brain are activated under some stimuli of interest. This approach permits controlling the Type I error at the level of clusters of adjacent voxels, gaining power with respect to voxel-wise inference approaches by exploiting the intrinsic spatial structure of fMRI data \citep{nichols2003controlling}.

However, the method is affected by the so-called spatial specificity paradox. This paradox arises because the larger the identified cluster, the less information we obtain from classic cluster inference about the signal within it. Indeed, the method tests the null hypothesis that none of the voxels in the cluster are active. Rejecting this null hypothesis only allows to claim the presence of at least one active voxel within the cluster. Consequently, larger clusters provide less information about the number and spatial location of active voxels \citep{woo2014cluster}. Moreover, conducting follow-up inference within the cluster, or ``drilling down,'' introduces a ``double-dipping'' problem and leads to an inflated Type I error rate \citep{kriegeskorte2009circular}.

The spatial specificity paradox can be resolved by making post-hoc inference on the True Discovery Proportion (TDP), i.e., the proportion of false null hypotheses within a subset. In neuroimaging, post-hoc TDP inference procedures provide lower confidence bounds on the proportion of active voxels within clusters, simultaneously over all possible clusters of interest. The simultaneity characteristic of the confidence bounds makes them valid even under post-hoc selection, allowing for follow-up inference within the cluster, unlike the cluster-extent-based thresholding approach \citep{rosenblatt2018all, goeman2023cluster}.

The first approach that proposed simultaneous inference on TDP in the fMRI context is the ``All-Resolution Inference'' (ARI) method developed by \cite{rosenblatt2018all}. However, ARI is parametric and can have low power in some scenarios, especially if correlated data such as fMRI are analyzed. It is well known that statistical analyses based on permutation theory are superior in terms of power and underlying assumptions in fMRI data analysis since they adapt to the correlation structure of the $p$-values \citep{winkler2014permutation, helwig2019statistical}. Permutation-based approaches to compute the lower bound of the TDP were first proposed by \cite{meinshausen2006false} and \cite{hemerik2019permutation}. However, these methods analyze only clusters consisting of the smallest $k$ $p$-values. The SansSouci method of \cite{blanchard2020post} 
extended this type of permutation-based simultaneous confidence bounds for the TDP to have the same flexibility as ARI, i.e., for clusters defined in different ways, even post-hoc, as many times as the researcher wants. An alternative permutation-based TDP method was proposed by \cite{vesely2023permutation}. 

Two recent approaches have appeared in the literature to compute a lower bound for the TDP: Notip by  \cite{blain2022notip} and pARI by \cite{andreella2023permutation}. Both methods build upon the work of \cite{blanchard2020post}, each proposing a different specific permutation-based TDP approach tailored to neuroimaging applications. In \cite{blain2022notip} work, the authors compare their methods with ARI and SansSouci; the gain in power and reliability of permutation-based approaches over parametric methods is apparent. However, due to the parallel publication process, Notip and pARI have not yet been compared to each other. \cite{blain2022notip} have made a comparison with pARI, but the settings of the method used in the study were not those recommended by \cite{andreella2023permutation}. Therefore, a proper comparative analysis is still lacking. In this manuscript, we provide such an analysis.

The paper is organized as follows. Section \ref{tdp} briefly revisits inference on the TDP. Subsection \ref{inferenceTDP} gives a general formulation of the permutation methods cited above (i.e., SansSouci, pARI, and Notip) before describing in detail the similarities and dissimilarities between Notip and pARI in Subsection \ref{sim}. Finally, Section \ref{neurovault} revisits the analyses presented in \cite{blain2022notip}, comparing them to pARI as defined in \cite{andreella2023permutation}. In this comparison, we follow \cite{blain2022notip} exactly in terms of the choice of the datasets and evaluation criteria. We show that we replicate the results shown in \cite{blain2022notip} regarding Notip, then add the pARI method under the specifications recommended by \cite{andreella2023permutation}. By following exactly the analysis choices made in the Notip paper, we make sure not to favor the pARI method, with which we are more familiar.

\section{Controlling True Discovery Proportions}\label{tdp}

Consider the brain $B= \{1, \dots, m\} \subset \mathbb{N}$ composed of $m$ voxels and, for each voxel $i\in B$, a $p$-value $p_i$ corresponding to the null hypothesis that it is not active under the condition of interest. We define by $A\subseteq B$ the unknown set of truly active voxels and by $S\subseteq B$ a generic non-empty subset of hypotheses of interest (i.e., a cluster of voxels). For any choice of $S$, interest lies in the number of true discoveries $a(S) = |A \cap S|$ or, equivalently, the TDP $|A \cap S|/|S|$, where $|S|$ stands for the cardinality of the set $S$. For a chosen error rate $\alpha\in (0,1)$, TDP procedures aim to construct lower $(1-\alpha)$-confidence bounds for these quantities, simultaneously over all possible choices of $S$. The confidence bounds for the number of true discoveries, denoted by $\bar{a}(S)$, are such that
\begin{equation}\label{eq:bounds}
    \Pr(\bar{a}(S) \le a(S)) \ge 1-\alpha
\end{equation}
for all $S \subseteq B$. An analogous formulation holds for the confidence bounds for the TDP, which can be immediately derived from $\bar{a}(S)$ \citep{goeman2011multiple}.

The simultaneity of the confidence bounds makes them valid even under post-hoc selection and so allows the user to decide which sets of hypotheses $S$ to analyze in a flexible and post-hoc manner. Therefore, methods with this property give information on the amount of true signal inside any set of voxels. The collection of voxels can be defined in various ways, allowing researchers to choose the method that suits their needs. Examples include clusters based on a searchlight, anatomical regions of interest (ROIs), functional ROIs, and data-driven regions (e.g., cluster-extent-based thresholding). Users can drill down into a region multiple times to more precisely identify the location of true active voxels by applying any region selection rule, whether data-driven or not.

\subsection{TDP based on critical vectors and permutations}\label{inferenceTDP}

To bound the TDP, pARI and Notip, like ARI and SansSouci, use a strategy based on critical vectors for ordered $p$-values. They compute the simultaneous lower $(1-\alpha)$-confidence bound for the number of true discoveries in a cluster $S$ as
\begin{equation}\label{eq:tdp}
    \bar{a}(S)=\max_{1\le u\le \mid S\mid} 1-u+| \left\{i\in S:{p}_i\le {\ell}_u\right\} | 
\end{equation}
where $\ell = (\ell_1, \dots, \ell_m) \in [0,1]^m$ is a suitable non-decreasing vector called critical vector, or in some cases template \citep{blain2022notip, blanchard2020post}. Different critical vectors have been proposed, but in order to obtain valid simultaneous confidence bounds as in Equation \eqref{eq:bounds}, it must satisfy the following condition:
\begin{equation}\label{eq:cond}
    \Pr\left(\bigcap_{i = 1}^{|N|} \{q_{(i)} \ge \ell_i\}\right) \ge 1-\alpha,
\end{equation}
where $N=B\setminus A$ is the unknown set of inactive voxels, and $q_{(1)}\leq\ldots\leq q_{(|N|)}$ are their sorted $p$-values. This means that the curve of the sorted $p$-values corresponding to inactive voxels should lie completely above the critical vector with probability at least $1-\alpha$.

\begin{figure}
\centering
\begin{minipage}{.45\textwidth}
  \centering
    \includegraphics[width = \linewidth]{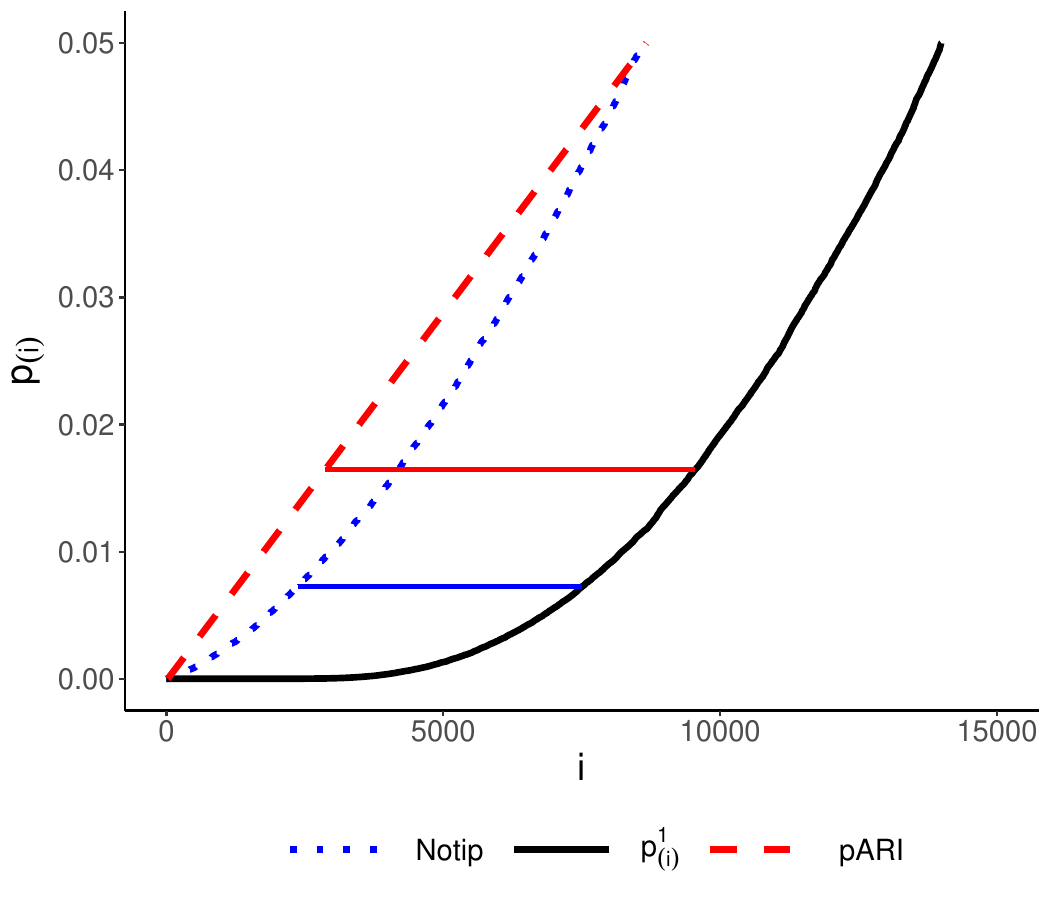}
    \caption{Graphical intuition of Equation \eqref{eq:tdp}. The black solid line represents the vector of sorted observed $p$-values $p_{(1)}\leq\ldots\leq p_{(m)}$. For each method (red for pARI, blue for Notip), the broken line represents the resulting critical vector; then $\bar{a}(S)$ is computed as the length of the solid segment, which is the largest distance between the curve of the observed $p$-values and the critical vector.}
    \label{fig:templates}
\end{minipage}%
\hfill
\begin{minipage}{.45\textwidth}
  \centering

        \includegraphics[width = \linewidth, height=5.2cm]{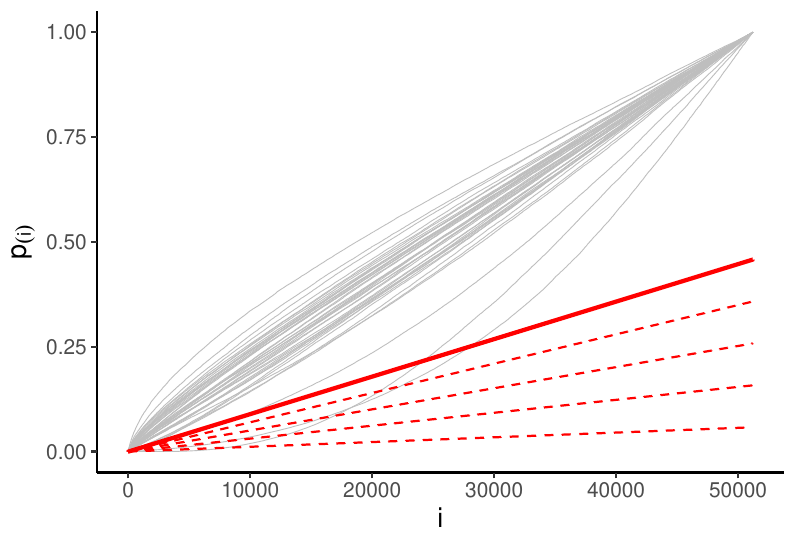}
    \caption{$\lambda_\alpha$-calibration step: the grey lines represent the vector of sorted $p$-values given by a random permutation of the data randomly sampling $40$ permutations. The red dashed lines are the candidate critical vectors for pARI having different $\lambda$ values. The solid red line is the optimal pARI critical vector having the largest $\lambda$ across the ones that cross the null distribution of the $p$-values represented by the grey lines at most $\alpha\%$ of the times.}
    \label{fig:null}
\end{minipage}
\end{figure}

In Figure \ref{fig:templates}, we give a graphical intuition of the computation of $\bar{a}(S)$, as defined in \eqref{eq:tdp}. The solid black line is the curve of the sorted $p$-values in the cluster $S$ of interest; the dashed red and dotted blue lines are two critical vectors (of pARI and Notip, respectively). If there were no signal in $S$, the 
black curve would be completely to the left of (i.e., above) each critical vector with probability $1-\alpha$. As it happens, the curve is way to the right of (i.e., below) the critical vector, indicating the presence of much signal. The lower bound $\bar{a}(S)$ to the number of active voxels, according to \eqref{eq:tdp}, is given as the maximal horizontal distance between the curve and the critical vector. It is clear from the figure that the shape of the critical vector is crucial and that different critical vectors may give very different TDP values.

To construct a critical vector that satisfies Equation \eqref{eq:cond}, both Notip and pARI rely on a high number $w$ of transformations of the data, $w-1$ of which can be random permutations or sign-flipping transformations or any other random data transformations that preserve the distribution of the test statistics under the null hypothesis \citep{winkler2014permutation}, while the remaining one must be the original, untransformed data \citep{hemerik2018exact}. The $p$-value curves arising from $w=40$ such data transformations are illustrated in Figure \ref{fig:null}, with each thin grey curve a $p$-value curve for a permutation. To find the critical vector, a pre-specified set of candidate critical vectors $\ell(\lambda)=(\ell_1(\lambda), \ldots, \ell_m(\lambda)$), $\lambda \in \Lambda$, is chosen, such that each $\ell_i$ is non-decreasing in $\lambda$. These candidate critical vectors are illustrated as the dashed red lines in Figure \ref{fig:null}. In order to satisfy \eqref{eq:cond}, the final critical vector is chosen as the highest curve such that $(1-\alpha)100\%$ of the sorted $p$-value curves lie above it. That is, 
if $p_{(1)}^j \leq \ldots \leq {p}_{(m)}^j$ are the sorted $p$-values obtained for the $j$-th random permutation, then $\lambda$ is chosen as the largest value such that
\begin{equation}\label{eq:calibration}
    |\{j: p_{(1)}^j > \ell_1(\lambda),\dots, {p}_{(m)}^j > \ell_1(\lambda)\}| \geq (1-\alpha)w.
\end{equation}
The resulting critical curve is given as the thick red line in Figure \ref{fig:null}.

This permutation-based process allows the method to incorporate the unknown spatial correlation structure of voxels in the calibration of the critical vector, and so to gain power compared to parametric methods.

\subsection{Differences between pARI and Notip}\label{sim}

The construction just described is common to pARI and Notip. However, pARI and Notip differ in their definition of the set of candidate vectors from which the optimal critical vector is selected, which we call a family of critical vectors (also called, in some cases, a set of learned templates as in \citet{blain2022notip} and \citet{blanchard2020post}).

For neuroimaging data, \cite{andreella2023permutation} recommend the shifted Simes family, given by
\begin{equation}\label{eq:template}
\ell_i (\lambda) = \dfrac{(i - \delta) \lambda}{m - \delta}
\end{equation}
where $\delta\in \{0,1,\ldots,m-1\}$, a shift parameter, is a fixed value that must be chosen independently of the data. The SansSouci approach used the same Simes-based family defined in Equation \eqref{eq:template} with $\delta=0$. Choosing $\delta$ larger has the result of losing all power for clusters $S$ of size $\delta$ or less, but in a trade-off, this results in substantially higher power for larger clusters. 
\cite{andreella2023permutation}, therefore, recommended $\delta>0$ in general, following \cite{hemerik2019permutation}, and substantially larger than $1$ if interest is in large clusters. However, $\delta$ is not allowed to depend on the sizes of clusters found, so a sensible default must be fixed. They recommended $\delta =3^3 = 27$ when interest is on clusters of large size, as is common in neuroimaging, so we take this as pARI's default value.

\cite{blain2022notip}, in contrast, define the family using $\tilde w$ permutations on external data with $\tilde m \approx m$ voxels. Let $\tilde p_{(1)}^j\leq \ldots\leq \tilde p_{(\tilde m)}^j$ be the sorted vector of $p$-values for the $j$-th permutation of the external data. In the family of candidate critical vectors proposed by \cite{blain2022notip}, $\ell_i(\lambda)$ is the $\lambda$-quantile of the vector $(\tilde p_{(i)}^1, \ldots, \tilde p_{(i)}^{\tilde w})$ if $i \leq k_{max}$, and $\ell_i(\lambda)=1$ otherwise, where $k_{\text{max}} \in \{1, \dots, m\}$ is some fixed bound chosen a priori. Formally,
\begin{equation}\label{eq:notip}
    \ell_i(\lambda) = \begin{cases} \tilde p_{(i)}^{(\lfloor\lambda \tilde w\rfloor)} 
    & i \leq k_{max} \\
  1 &  \text{otherwise,}
\end{cases}
\end{equation}
where $\tilde{p}_{(i)}^{(j)}$ denotes the $j$-th smallest value among $\tilde p_{(i)}^1, \ldots, \tilde p_{(i)}^{\tilde w}$.

Though seemingly similar in their use of permuted data, Equation \eqref{eq:notip} is markedly different from \eqref{eq:calibration} above since \eqref{eq:notip} uses only the marginal distribution of the ordered $p$-values, whereas \eqref{eq:calibration} uses their joint distribution. The relationship between the external data and the data under analysis should, therefore, not be seen as the usual relationship between a training and a validation set. In fact, \cite{meinshausen2006false} proposed using the same data in \eqref{eq:calibration} and \eqref{eq:notip}, and though \cite{hemerik2019permutation} and \cite{blanchard2020post} pointed out that doing so destroys the formal validity of the method, the choice of \cite{meinshausen2006false} is generally fine in practice.

In Notip, $k_{\text{max}}$ is a tuning parameter, compable to $\delta$ in pARI, and like $\delta>0$, use of $k_{max}< m$ was recommended for a different family by \cite{hemerik2019permutation}. Effectively, all $p$-values higher than the $k_{max}$-th one are ignored by Notip. Like $\delta$, the choice of $k_{\text{max}}$ induces a trade-off: small values can lead to a less conservative family of critical vectors but also to smaller lower bounds for the TDP. \cite{blain2022notip} describe $k_{\text{max}}$ as the largest size of the cluster for which a high proportion of active voxels is guaranteed.  They suggested to fix $k_{\text{max}} = 1000$. 

As a further improvement, \cite{andreella2023permutation} proposed a step-down version of pARI, which outperforms the SansSouci method in terms of power even if the same critical vector family is used. This improvement comes at the price, however, of high computational time. In this paper, we use the faster version of pARI without the step-down.

\section{Comparison on Neurovault data}\label{neurovault}

In this section, we compare the Notip and pARI approaches, following exactly the analysis performed originally by \cite{blain2022notip}. The Neurovault database \citep{varoquaux2018atlases} contains data from many fMRI studies. Here, we analyzed collection $1952$ (\url{http://neurovault.org/collections/1952}), consisting of statistical maps from $20$ different studies. The data were preprocessed using the Python code made available by \cite{blain2022notip} at \url{https://github.com/alexblnn/Notip}, finally having $36$ contrast pairs. The analysis was carried out using the \texttt{pARI} \texttt{R} package (\url{https://CRAN.R-project.org/package=pARI}) for applying pARI, and the Python code made available by \cite{blain2022notip} for applying Notip. Figures \ref{fig:templates} and \ref{fig:null}, above, have been computed using the first dataset of this collection, i.e., ``shapes versus baseline'' contrast versus ``faces versus baseline'' contrast from the HCP study. To make Figure \ref{fig:templates} clearer, we considered the cluster composed of the smallest $15{,}000$ voxels.

We will redo only those analyses from \cite{blain2022notip} in which they compare performance between the Notip and competing methods. It is not straightforward to compare different TDP methods because each method gives $2^m$ TDP confidence bounds. A method that performs better for some TDP bounds may be worse for other bounds, even within the same data or simulation scenario. We follow \cite{blain2022notip} in their choice of metric for comparing methods, which focuses on the size of the largest cluster found at a fixed TDP threshold. Other metrics are possible; e.g., \cite{andreella2023permutation} used the TDP of clusters defined at a fixed cluster-defining threshold as their metric. In all the analysis, we fix the number of permutations used to compute the Notip critical vector $\tilde{w}$ to $10000$, and the number of permutations used to calculate the null distribution of the $p$-values to $1000$.

\begin{figure}
    \centering
    \includegraphics[width = .5\textwidth]{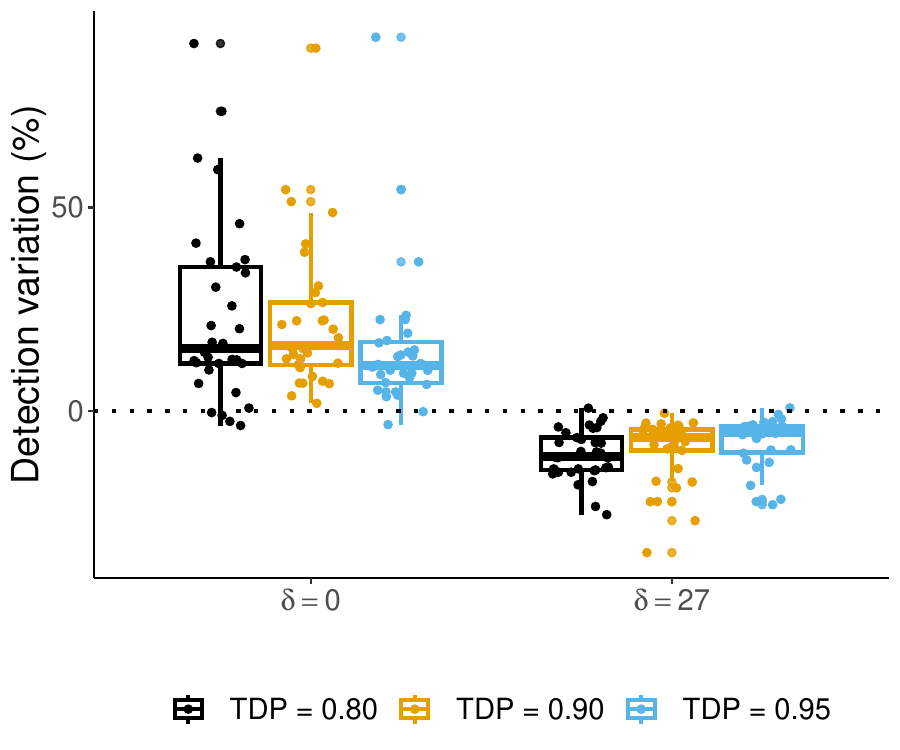}
    \caption{Percentage variations detected defined as $ \dfrac{|S|_{\text{Notip}} - |S|_{\text{pARI}}}{|S|_{\text{pARI}}}$. The left side is the non-recommended setting for pARI (i.e., fixing $\delta = 0$), which we show only to reproduce the results of \cite{blain2022notip}. Instead, the right side represents the results using the recommended setting for pARI as shown by \cite{andreella2023permutation} when $\delta = 27$. Since the comparison is given in terms of variation as defined above, values below $0$ indicate better performance in pARI than in Notip.}
    \label{fig:boxplot}
\end{figure}

The left-hand side of Figure \ref{fig:boxplot} reproduces the results of \citet[Figure 4, right-hand side]{blain2022notip}, in which they compare Notip to pARI with $\delta=0$, i.e., to SansSouci. The relative number of detections between Notip and pARI, defined as
\begin{equation}\label{var}
    \dfrac{|S|_{\text{Notip}} - |S|_{\text{pARI}}}{|S|_{\text{pARI}}},
\end{equation}
where $|S|$ is the largest possible region that reaches a fixed TDP level, is analyzed. The boxplots presented in Figure \ref{fig:boxplot} show the distribution of this metric over $36$ contrasts maps from Neurovault collection $1952$ data and TDP thresholds $0.8$, $0.9$, $0.95$ with $\alpha$ fixed at $0.05$. The results on the left-hand side of Figure \ref{fig:boxplot} reproduce almost exactly the results presented in \cite{blain2022notip}. There are minor differences due to the use of random permutations. In addition, we noticed that the code provided by \cite{blain2022notip} did not consider the mandatory inclusion of the identity transformation, which we included to get exact $\alpha$ control \citep{hemerik2018exact}, even though due to the high number of permutations (i.e., $w = 1000$) this make almost no difference. The right-hand side of Figure \ref{fig:boxplot} makes the same comparison, but with pARI's recommended setting of $\delta=27$.

\begin{figure}
    \centering
    \includegraphics[width = .6\textwidth]{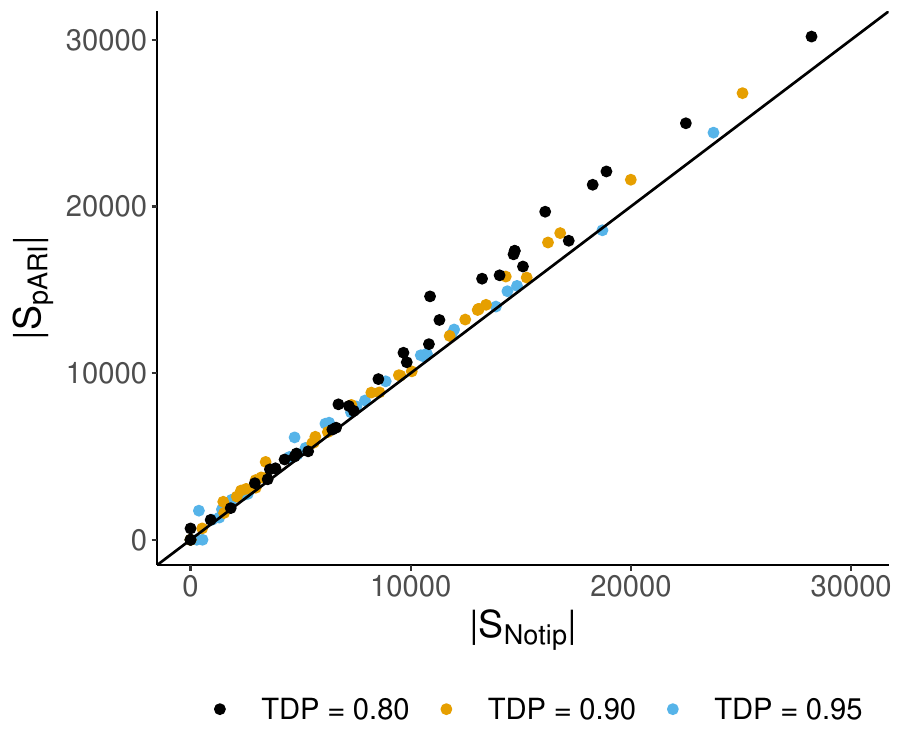}
    \caption{Size of the largest clusters found by pARI with $\delta = 27$ ($|S_{\text{pARI}}|$) and Notip ($|S_{\text{Notip}}|$) with TDP $\ge t \in \{0.8, 0.9, 0.95\}$.}
    \label{fig:scatterplot}
\end{figure}

Where Notip almost always outperformed pARI without the shift, we note that the reverse is true for the recommended shifted version of pARI. To investigate further, Figure \ref{fig:scatterplot} plots the largest cluster sizes found by pARI ($\delta=27$) against those found by Notip. Also, from this plot, we see that the size of the largest cluster found is almost always greater with pARI than with Notip, and this effect is especially pronounced when the largest cluster contains many voxels (i.e., top right part of Figure \ref{fig:scatterplot}). 

Finally, Table \ref{tab:results} reproduces results from Table 2 in \cite{blain2022notip}, to which we added results for pARI with $\delta = 27$. The contrast pair ``look negative cue vs look negative rating'' of the Neurovault database is analyzed. The clusters are computed by thresholding the statistical map at absolute values greater than $3$ and keeping only clusters composed of at least $150$ voxels \citep{woo2014cluster}. Again, we can note how imposing $\delta = 27$ significantly increases the method's power; pARI is, in fact, more powerful than Notip in all clusters, except the smallest one, i.e., it returns greater lower bounds for the TDP. 

\begin{table}[h]
    \centering
\begin{tabular}{rr|r|rr}
\toprule
& & \multicolumn{3}{c}{True Discovery Proportion} \\
\cmidrule(r){3-5}
& & Notip &\multicolumn{2}{c}{Simes-based pARI} \\
Cluster ID  & Cluster Size  &  & $\delta = 0$ & $\delta = 27$\\
\midrule
1 & 7,695 & 0.26 & 0.23 &  \textbf{0.34}\\
2 & 14,877 & 0.45& 0.32 &  \textbf{0.58}\\
3 & 14,445& 0.50& 0.37  &\textbf{0.60}\\
4 & 5,238 &0.29 & 0.24  &\textbf{0.34}\\
5 & 4,563 & \textbf{0.30}& \textbf{0.30}  &0.29\\
6 & 12,555 &0.35 & 0.16  &\textbf{0.52}\\
7 & 6,075 & 0.17& 0.09  &\textbf{0.24}\\
8 & 25,812 &0.66 &  0.46 &\textbf{0.76}\\
9 & 6,507 &0.17 & 0.15  &\textbf{0.20}\\
\bottomrule
\end{tabular}
    \caption{Clusters identified with threshold $|z| >3$: clusters size and TDP lower bound at risk level $\alpha = 0.05$ using two possible critical vectors (Notip, and Simes-based pARI with $\delta=27$) on contrast pair ``look negative cue vs look negative rating.'' For each cluster, the values in bold indicate the best result, i.e., TDP (lower limit) higher.}
    \label{tab:results}
\end{table}

We can conclude that the shifted version of Simes-based pARI performs remarkably well and, in most cases, surpasses the Notip approach, emphasizing the importance of choosing an appropriate critical vector (and shift value) for gaining power.

\section{Discussion}

We have seen that pARI outperformed Notip in almost all settings considered by \cite{blain2022notip} when the shift parameter $\delta$ of pARI was appropriately set at $\delta=27$. This finding may seem counterintuitive since Notip uses additional information in the form of external data. It should be realized, however, that in this external data, Notip looks only marginally at the ordered $p$-values. The added value of this information may be limited in practice, as also illustrated by the experience \citep{meinshausen2006false, blain2022notip} that double dipping by reusing the data under analysis as if they were external does not break the validity of the method in practice.

Both Notip and pARI have a tuning parameter ( $k_{max}$ and $\delta$, respectively). The presence of an additional parameter can be considered a drawback, especially since it has to be chosen before seeing the data. Both methods, therefore, recommend a default value ($k_{max}=1000$ and $\delta=27$) for applications in neuroimaging. It is interesting to note that $k_{max}$ and $\delta$ have complementary effects: $k_{max}<m$ focuses power of Notip away from very large clusters, while $\delta>0$ focuses power of pARI away from small ones. It could be an interesting avenue of further research to formulate an alternative method that has both a $k_{max}$ and a $\delta$ parameter \citep[e.g., as considered in a different context by][]{hemerik2019permutation}.

It can be argued that Notip has a second tuning parameter in the choice of the external data. This can be avoided by re-use of the data under analysis, but the resulting method has no formal proof of error control. Whether data are reused or not, this additional analysis step makes the procedure more computationally expensive. For the analyses presented here (i.e., considering standard Notip and the single-step version of pARI), Notip takes approximately $42$ minutes, while pARI takes only $1$ minute. pARI, on the other hand, becomes computationally expensive if the step-down version is used.

Various trade-offs characterize both methods and can be seen as two out of many possible analysis choices. The comparison that we have given here shows that the choice of the family matters, but further analyses are needed to study each method's power properties in more detail and to determine which method should be preferred in which settings. This could also help in finding even better families than those considered by Notip and pARI.

\section*{Ethics}
This research relies on existing data sources, and no primary data collection was undertaken.

\section*{Data and code availability}
The data underlying this study are those used in \citet{blain2022notip}, available in the NeuroVault database at \url{http://neurovault.org/collections/1952}. The code to preprocess the data and apply the Notip method is available at the GitHub repository \url{https://github.com/alexblnn/Notip}. The code for the pARI method is developed in the \texttt{R} package \texttt{pARI}, at \url{https://CRAN.R-project.org/package=pARI}.

\section*{Authors contribution}
\textbf{Angela Andreella}: conceptualization, methodology, software, data curation, formal analysis, investigation, writing - original draft, writing - review \& editing. \textbf{Anna Vesely}: conceptualization, methodology, data curation, formal analysis, writing - review \& editing. \textbf{Wouter Weeda}: conceptualization, methodology, writing - review \& editing, supervision. \textbf{Jelle Goeman}: conceptualization, methodology, writing - review \& editing, supervision.

\section*{Declaration of competing interest}
The authors declare no competing interests.

\section*{Acknowledgements}
Angela Andreella gratefully acknowledges financial support from Ca' Foscari University of Venice via Grant No.~PON 2014-2020/DM 1062. Anna Vesely acknowledges financial support by the Deutsche Forschungsgemeinschaft (DFG) via Grant No.~DI 1723/5-3. The authors express gratitude to Jesse Hemerik for valuable discussions related to this work.

\section*{Appendix}
\allowbreak
The pARI approach proposed by \cite{andreella2023permutation} depends on the choice of the parameter $\delta$ that impacts directly the inference power.
\cite{andreella2023permutation}, after analyzing several fMRI datasets with different signal-to-noise structures, suggests the following settings: a shift of at least $1$, in general, and a larger shift, set as a default at $27$, if clusters composed of many voxels are of interest, as is usual in neuroimaging. In this Appendix, we revisit these settings to see if the choice of $27$ is indeed close to optimal in the data sets considered by \cite{blain2022notip}.

\begin{figure}
    \centering
    \includegraphics[width = .5\textwidth]{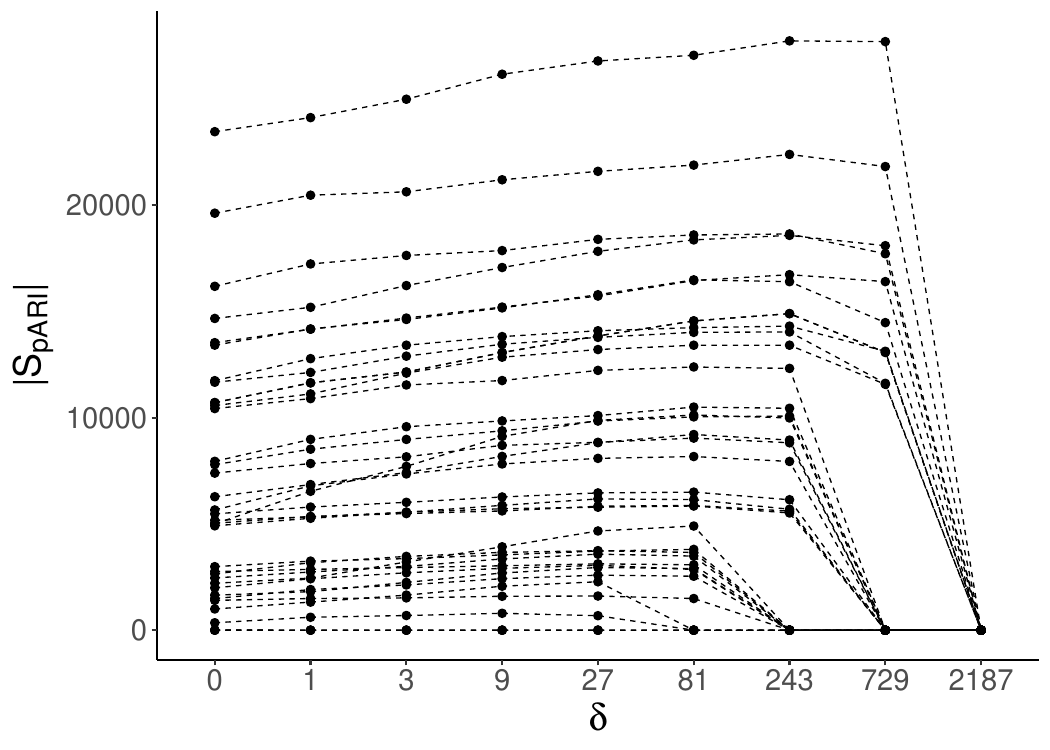}
    \caption{Largest size founded by pARI considering $\delta \in \{0,1,3,9,27, 81, 243, 729, 2187\}$ for each pair of the $36$ Neurovault contrasts that reaches at least a lower bound for the TDP equals $0.9$.}
    \label{fig:deltas}
\end{figure}

Figure \ref{fig:deltas} shows the size of the largest cluster detected by pARI, considering several values for the shift parameter, i.e., $\delta \in \{0,1,3,9,27, 81, 243, 729, 2187\}$ analyzing the $36$ pairs of contrasts from the Neurovault data, collection 1952 \citep{varoquaux2018atlases} fixing TDP $\ge 0.9$. The optimal $\delta$ value generally depends on the analyzed dataset. However, we can note how in the $36$ pairs of contrasts analyzed, cluster size tends to increase with $\delta$ initially until it drops towards zero when $\delta$ approaches the cluster size. Since all clusters of interest are substantially larger than $\delta = 27$, this seems a reasonable choice, as found using different arguments by \cite{andreella2023permutation}. 

Table \ref{tab:results2} extends the results presented in Table \ref{tab:results} by showing the lower bounds for the TDP computed by the pARI approach with $\delta \in \{0,1,3,9,27\}$. We can note that there already for $\delta=1$ the results are competitive with Notip. 
Results appear to be, to an extent, fairly robust against the choice of $\delta$, as long as a positive value is considered that is substantially smaller than the size of the smallest cluster of potential interest. 

\begin{table}[h]
    \centering
\begin{tabular}{rr|r|rrrrr}
\toprule
& & \multicolumn{6}{c}{True Discovery Proportion} \\
\cmidrule(r){3-8}
& & Notip &\multicolumn{5}{c}{Simes-based pARI} \\
Cluster-ID  & Cluster Size  &  & $\delta = 0$ & $\delta = 1$ & $\delta = 3$ & $\delta = 9$ & $\delta = 27$\\
\midrule
1 & 7,695 & 0.26 & 0.23 & 0.28 & 0.33 & \textbf{0.36} & 0.34\\
2 & 14,877 & 0.45& 0.32 & 0.41 & 0.49 & 0.55& \textbf{0.58}\\
3 & 14,445& 0.50& 0.37 &0.46 & 0.55 & 0.59 &\textbf{0.60}\\
4 & 5,238 &0.29 & 0.24 &0.31 & 0.36 & \textbf{0.38} &0.34\\
5 & 4,563 & 0.30& 0.30 &0.33 & \textbf{0.37} & 0.36 &0.29\\
6 & 12,555 &0.35 & 0.16 & 0.29 & 0.40 & 0.48 &\textbf{0.52}\\
7 & 6,075 & 0.17& 0.09 &0.17 & 0.24 & \textbf{0.27} &0.24\\
8 & 25,812 &0.66 &  0.46&0.59& 0.67 & 0.73 &\textbf{0.76}\\
9 & 6,507 &0.17 & 0.15 &0.19 & 0.22 & \textbf{0.23} &0.20\\
\bottomrule
\end{tabular}
    \caption{Clusters identified with threshold $z >3$: clusters size and TDP lower bound at risk level $\alpha = 0.05$ using six possible critical vectors (Notip, Simes-based pARI with $\delta\in\{0,1,3,9,27\}$) on contrast pair ``look negative cue vs look negative rating''. For each cluster, the values in bold indicate the best result, i.e., TDP (lower limit) higher.}
    \label{tab:results2}
\end{table}

\newpage

\bibliographystyle{apalike} 
\bibliography{bibliography.bib}

\end{document}